\documentclass{ws-mpla}

\begin{document}

\markboth{A.Molochkov}{Sea-quark flavor asymmetry 
in the nucleon 
from a relativistic analysis
of the Drell-Yan scattering off nuclei}

%%%%%%%%%%%%%%%%%%%%% Publisher's Area please ignore %%%%%%%%%%%%%%
\catchline{}{}{}{}{}
%%%%%%%%%%%%%%%%%%%%%%%%%%%%%%%%%%%%%%%%%%%%%%%%%%%%%%%%%%%%%%%%%%%

\title{Sea-quark flavor asymmetry 
in the nucleon 
from a relativistic analysis 
of the Drell-Yan scattering off nuclei}

\author{\footnotesize ALEXANDER MOLOCHKOV}

\address{Physical Department, Far Eastern National University, Sukhanova st., 8\\
Vladivostok, 690950,
Russia\\
alexm@ifit.phys.dvgu.ru}

\maketitle

\pub{Received (Day Month Year)}{Revised (Day Month Year)}

\begin{abstract}
The study presented in this paper shows that accounting for the relativistic structure
of the deuteron allows to explain the ratio of the Drell-Yan pair production cross-section 
at the low Bjorken $x$ off the deuteron
and the proton. 
Thus, the sea quark distributions in the nucleon should be studied with
accounting for the effects of the relativistic structure of the deuteron.
The suggested approach reduces theoretical uncertainty in extracting
the ratio $\bar u/\bar d$ from the data and it is important for the
clarification of the nature of the sea quark asymmetry in the nucleon.
\keywords{Drell-Yan; parton distribution; Bethe-Salpeter.}
\end{abstract}

\ccode{PACS Nos.: 25.30.Mr, 03.65.-w, 11.10.St, 13.60.Hb, 11.80.-m}

\section{Introduction}	

The Drell-Yan pair production remains up to now the theoretically cleanest way
to access sea-quark distributions in
hadrons. 
Necessary stage of such analysis is the measurement both with
protons and neutrons, in which  sea-quarks with different flavors
are probed.
As the construction of a target containing free neutrons is not feasible,
nuclear targets have to be employed. In the experiment that
used the simplest nuclear target --- the deuteron,  the flavor asymmetry
of the sea-quark distributions in the nucleon has been observed~\cite{DYDeut}.

At the same time, analysis of the deep inelastic scattering off different nuclei
shows that nucleon valence quark structure changes beyond
mass-shell~\cite{aub83}.
Recent studies within the Bethe-Salpeter formalism point to the relativistic nature
of this phenomenon.
Relativistic effects should be manifested as modification of the
sea quark distribution in the bound nucleon and, therefore, will
 change the contribution of the sea quark component in the
nucleon structure extracted from the deuteron data. This, in turn, 
affects estimations of the anti-quark component made by analyzing
the deuteron and proton data.

To clarify the role of the relativistic effects in the Drell-Yan (DY)
pair production, we study this process within the Bethe-Salpeter approach.

\section{Drell-Yan Crossection}

Since the amplitude for the Drell-Yan scattering is defined by the imaginary part
of the amplitude for the
forward scattering $\langle A|T(J_{\mu} J_{\nu})|A \rangle$ in the cross-channel, 
we can obtain it within the Bethe-Salpeter formalism developed for DIS~\cite{approach}. 
Within this approach the matrix element is defined
by the nucleon space-time distribution and
vacuum average of the T-product of the nucleon fields and
nucleon electromagnetic current~( see Ref.~\cite{approach}).
Following this method we get the following expression of the cross-section for the 
Drell-Yan scattering off deuteron:
\begin{equation}
\sigma^{\rm pD}(P,q)= \int\frac{d^4p}{(2\pi)^4}\frac{\sigma^{\rm
pN}\left(p,q\right) f^{{\rm
N}/D}(P,p)}{\left(p^2-m^2\right)^2
(\left(P-p\right)^2-m^2)}. \label{fermi4d}\end{equation}
This expression gives the cross-section of the DY pair production off the deuteron
in terms of the off-mass-shell DY nucleon cross-section $\sigma^{\rm pN}$ and the
nucleon distribution function $f^{\rm N/D}(P,p)$. This
distribution function is defined by the
 Bethe-Salpeter amplitudes, and together with the denominator it composes
 the four-dimensional momentum distribution of the struck nucleon
 inside the deuteron carrying the total momentum
 $P=(M_{\rm D},{\bf P})$.
In that way Eq.~(\ref{fermi4d}) expresses the nucleon blurring that results
from the four-dimensional distribution of the nucleon inside the nucleus.
By analogy with the non-relativistic $3$D momentum distribution we call it
four-dimensional Fermi motion.

Due to the four-dimensional integration in Eq.~(\ref{fermi4d}) actual
calculations require information about nucleon cross-section $\sigma^{\rm pN}$
in the kinematical region of the off-mass-shell values of the
nucleon energy $p_0$ ($p_0^2 \neq {\bf p}^2 + m^2$).
The off-mass-shell behavior of $\sigma^{\rm pN}$ is unobservable, since then
explicit microscopic calculations of the cross-section have no experimental
reference and strongly model dependent.
Thus, Eq.~(\ref{fermi4d}) has to be rewritten in terms of measurable
quantities such as the cross-section for DY scattering off the physical nucleon.
The simplest solution of this problem is provided by
the integration in Eq.~(\ref{fermi4d})
with respect to $p_0$.
Analytical properties of the integrand in Eq.~(\ref{fermi4d}) give a
way to do it explicitly.
Within the assumption that $\sigma^{\rm pN}(p,q)$ and $f^{N}(P,p)$ are regular
with respect to $p_0$
the integrand  contains a second order pole corresponding to the struck
nucleon and a first order pole corresponding to the spectator.
These poles lie in the different
half-planes of the complex plane $p_0$. So, we can choose one of the
singularities to perform the integration. To express the
nuclear hadron tensor in terms of the nucleon structure functions
it is necessary to choose the second
order pole ($p_0-E_{\rm N}$) that corresponds to the struck nucleon.
The result of the contour integration in vicinity of the second order pole in
Eq.(\ref{fermi4d}) is defined by the derivative of the pole residue
with respect to $p_0$ at the point $p_0=E_{\rm N}$.

Doing the integration and using the relation in the Bjorken limit 
$\sigma(P_{\rm p},P_{\rm D},q)=\sigma(x_1,x_2)$
we get the following expression for the cross section of the DY pair production off the deuteron:

\begin{equation}
\sigma^{\rm pD}(x_1,x^{\rm D}_2)=
\int\frac{d^3p}{(2\pi)^3}
\left[\sigma^{\rm N}(x_1,x_2^{\rm N})- \frac{{\Delta^{\rm N}_{\rm D}}}{E_{\rm N}}
x_2^{\rm N}\frac{d \sigma^{\rm pN}(x_1,x^{\rm N}_2)} {dx^{\rm N}_2} \right]
\frac{f^{N/D}(M_D,{\bf p})}{8M_{D}E_{\rm N}^2{\Delta^{\rm N}_{\rm D}}^2},\label{S2A}
\end{equation}
where
$x_2^{\rm N}=x_2^{\rm D}m/(E_{\rm N}-p_3)$ is the Bjorken $x$ of the struck nucleon,
$E_{\rm N}=\sqrt{m^2+{\bf p}^2}$ is the nucleon on-shell energy,
$\Delta^{\rm N}_{\rm D}=M_{\rm D}-2E_{\rm N}$ 
is equivalent to the total energy shift of the struck nucleon
due to the binding and Fermi motion.

 The function $f^{\rm N/D}(M_{\rm D},{\bf p})$ together with the denominator composes the
 three dimensional momentum
distribution of the struck nucleon inside the deuteron.
According to the normalization condition
for the Bethe-Salpeter vertex function this distribution satisfies the baryon
 and momentum sum rules~\cite{mynucl} and coincides with the usual nuclear
momentum distribution. Thus, the first term in Eq.~(\ref{S2A}), which results from the
derivative of the propagator of the nuclear spectator,
expresses contribution of the conventional $3$D-Fermi motion.

The second term with $d\sigma^{\rm pN}({x_2}_{\rm N})/d{x_2}_{\rm N}$ results from the
nucleon DIS amplitude derivative:
$$\frac{d\sigma^{\rm pN}_{\mu\nu}(p,q)}{dp_0}=
\frac{dx_2^{\rm N}}{dp_0}\frac{d\sigma^{\rm pN}(x_1,x_2^{\rm N})}{dx_2^{\rm N}}$$
This term is responsible for 
the deviation from unity of the ratio of the cross sections. 
Its contribution is proportional to the coefficient $\Delta^{\rm N}_{\rm D}$;
which, therefore,
characterizes contribution of the four-dimensional Fermi-motion.

It is important to note that the struck nucleon pole gives the
factor $1/{\Delta^{\rm N}_{\rm D}}^2$, since then the contribution
from all other singularities (for example nucleon self-energy cut
or anti-nucleon pole) are suppressed at least as $(\Delta^{\rm
N}_{\rm D}/M_{\rm D})^2$~\cite{approach}. Since the mean value of the energy
shift is small ($\Delta^{\rm N}_{\rm D}/M_{'rm D} \propto
O(10^{-2})$), Eq.(\ref{S2A}) provide deuteron Drell-Yan cross-section
with accuracy up to terms of order $(\Delta^{\rm
N}_{\rm D}/M_{\rm D})^2\propto O(10^{-4})$.

\section{Numerical results}

\begin{figure}[ph]
\centerline{\psfig{file=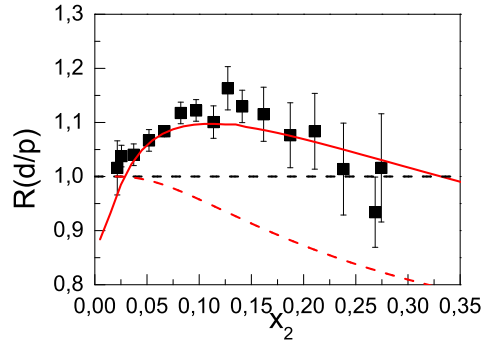,width=2.5in}}
\vspace*{8pt}
\caption{Calculations of the DY cross-section ratio for the deuteron and proton
compared with FNAL E866 (2001) data. 
The dashed curve is the calculation within the approximation $\sigma^{\rm pD}=(\sigma^{\rm pp}+\sigma^{\rm pn})/2$, 
the solid curve is the calculation with the four-dimensional Fermi-motion 
taken into account. Both of the calculation use the GRV parameterization with 
sea-quark flavour symmetry preserved ($\bar u/\bar d=1$).\protect\label{fig1} }
\end{figure}

Using the Gluck-Reya-Voght (GRV) parameterization~\cite{GRV} for the $\sigma^{\rm pN}$ and solution of
the Bethe-Salpeter equation with the Graz-II separable kernel~\cite{graz} as $f^{\rm N/D}(M_{\rm D},{\bf p})$ 
we make
numerical calculation fo the ratio $\sigma^{\rm pD}/\sigma^{\rm pp}$.
It is important to note that the full GRV parameterization is obtained within
the approximation $\sigma^{\rm pD}=(\sigma^{\rm pp}+\sigma^{\rm pn})/2$, therefore it already 
implies isospin asymmetry of the sea-quark distribution in the nucleon in order to describe the DY data. 
To check influence of the relativistic nuclear
effects on the deviations of the ratio $\sigma^{\rm pD}/\sigma^{\rm pp}$ from unity
we use the GRV parameterization without quark-sea
asymmetry ($\bar u/\bar d=1$). The results of the calculation are presented in
Fig.\ref{fig1}. The dashed curve represents the calculation 
within the approximation with $\sigma^{\rm pD}=(\sigma^{\rm pp}+\sigma^{\rm pn})/2$, the
solid curve represent calculation with the four-dimensional Fermi-motion 
taken into account. 

It is clear from the comparison with the new FNAL data~\cite{FNAL},
that the relativistic kinematics can reproduce the deviation of the cross-section 
ration without the flavor asymmetry of the nucleon sea. 
It is important to note, that the $A$-dependence of the deviation is defined 
by the coefficient $\Delta^{\rm N}_{A}/M_{A}$ in front 
of the cross-section derivative same as in case of DIS. 

Thus, we can conclude that observed flavor asymmetry results from the 
relativistic kinematics of bound nucleons;
the bound nucleon structure change have the same nature in the DIS and DY processes; 
the effect has the same $A$-dependence as EMC-effect.
A more detailed study of the $A$-dependence of the effect and high-$x$ data 
can shed more light on the nature of the sea-quark flavour asymmetry in the nucleon.

\section*{Acknowledgments}

I would like to thank  G.~Smirnov, U.~Mosel, H.~Toki and O.Lynnik for useful discussions. 
I also thank Research Center for Nuclear Physics of the Osaka University 
for the support and warm hospitality.
The work is made under partial support of the FENU Innovation Educational Programme in the framework 
of the National Project "Education".

\end{document}